\documentclass[prb,aps,showpacs,floatfix,twocolumn]{revtex4}
\usepackage{times}
\usepackage{amsmath,bm,amsfonts}
\usepackage{graphicx}
\usepackage{color}

\begin{document}

\title{Breakdown of half-metallic ferromagnetism in zinc-blende II-V compounds}

\author{Yun \surname{Li}}
\author{Jaejun \surname{Yu}}
\email[Corresponding author. Electronic address: ]{jyu@snu.ac.kr}
\affiliation{FPRD and Center for Theoretical Physics, Department of
  Physics and Astronomy, Seoul National University, Seoul 151-747, Korea}

\begin{abstract}
  We investigated the electronic and magnetic properties of a series of
  zinc-blend II-V compounds by carrying out density-functional-theory
  calculations including spin-orbit couplings. Contrary to the case of CaN
  and CaP, the half-metallic characteristics of the II-V compounds such as
  CaSb and CaBi were found to be destroyed.  Our analysis of the valence
  band structures of CaAs, CaSb, and CaBi revealed a critical role of the
  spin-orbit coupling interactions on the exchange-split band structure,
  thereby leading to breakdown of the half-metallic ferromagnetism for the
  systems with heavier group V elements in the zinc-blend II-V compounds.
\end{abstract}


\maketitle


Half-metallic ferromagnets (HMFs), being metallic in only one of the two
spin channels, have been considered as an indispensable ingredient in the
development of spintronic devices and applications. The realization of the
half-metallicity has been investigated extensively since its first
prediction on Heusler alloys by de Groot \textit{et al.}\cite{deGroot}.
However, due to the complexity in their electronic characteristics,
manifesting both metallic and insulating properties in a single system at
the microscopic level, HMF materials often require complex structures such
as ternary spinel, Heusler, and double perovskite structures, also
including transition-metal elements as a source of local magnetic moments.

To explore possible spintronics applications of HMFs to semiconductor
devices, there have been a great deal of studies exploiting the half
metallicity in the binary compounds of zinc-blende (ZB)
structure\cite{TM-Zhao, TM-Aki, TM-Miz, TM-Ono, TM-Xu, TM-Xie, TM-Gal,
  TM-San, II-Kus, TM-Yao, II-Sie, II-Yao}, which is simple and compatible
with existing III-V and II-VI semiconductors.  So far there have been a
few reports on the fabrication of nano-scale MnAs dots on GaAs, CrSb
ultra-thin films on the GaSb substrate, and ultra-thin CrAs layers in the
CrAs/GaAs multilayers\cite{TM-Zhao,TM-Aki,TM-Miz,TM-Ono}. In addition, the
half-metallic ZB compounds with transition-metal elements were also
reported by first-principles density-functional-theory (DFT) calculations,
\cite{TM-Gal} whereas a high Curie temperature above 400 K has been
observed in experiments for CrAs and CrSb systems\cite{TM-Aki,TM-Zhao}.

Contrary to the ZB compounds containing transition-metal elements, where
the localized $d$ electrons are responsible for the ferromagnetic
component in HMF, several DFT calculations have predicted another kind of
ZB II-V compounds as a candidate for the HMFs \emph{without} containing
any transition-metal element\cite{II-Kus, II-Sie, II-Yao}. The ZB
compounds of alkaline earth elements Ca, Sr, and Ba combined with all the
elements of the group V were shown to be HMFs with a magnetic moment of
one Bohr magneton ($\mu_{B}$) per formula unit-cell (f.u.). In these
compounds, the presence of a flat $p$ band crossing the Fermi level
($E_{\mathrm{F}}$) in its paramagnetic phase is a key to the half-metallic
electronic structure. The narrow $p$ band contributes to the exchange
energy splitting close to 0.5 eV, consequently leading to an insulating
gap in the majority spin channel and a metallic state in the minority spin
channel.

However, since the $p$ states of the heavy elements like Sb and Bi are
affected by the relativistic spin-orbit coupling (SOC), the SOC is
expected to play a role in the determination of the II-V valence band
structure consisting of the anion $p$ components of As, Sb, and Bi
atoms. Therefore, their ground state electronic and magnetic properties
needs to be examined carefully in connection with the half-metallicity of
the II-V compounds with heavy group V elements.

In this paper, we report the results of our non-collinear DFT calculations
including the spin-orbit coupling terms for the electronic and magnetic
properties of the ZB II-V compounds of $AX$ ($A$=Ca, Sr, and Ba; $X$=N, P,
As, Sb, and Bi).  Contrary to the previous reports on the ZB II-V
compounds, we found that the ZB II-V compounds containing Sb or Bi are no
more half-metallic due to the strong SOC overriding the exchange
instability triggered by the flat band feature of the anion
$p$ states. Our calculations showed a broad spectrum in the physical
properties of the ZB II-V compounds, ranging from the half-metallic
ferromagnet to an ordinary ferromagnetic metal to a paramagnetic
metal. Further analysis of the valence band structures explains the
competition between spin-orbit couplings and exchange interactions as a
key ingredient in the determination of ground states of the ZB II-V
compounds.


To investigate the role of spin-orbit coupling interactions, we
carried out non-collinear DFT calculations including the SOC to
determine the electronic structure and magnetic properties of the
ZB II-V compounds of $AX$ ($A$=Ca, Sr, and Ba; $X$=N, P, As, Sb,
and Bi). For the self-consistent electronic structure
calculations, we have optimized the equilibrium lattice constant
of each $AX$ compound under the symmetry constraint of the
zinc-blende (ZB) structure. The structural stability has been
extensively examined in the previous work\cite{II-Sie}. It was
discussed that the ZB-type structure of CaAs can exist as a
metastable state under certain conditions even though neither ZB
nor wurtzite (WZ) structures are the absolute ground state of
CaAs. Without further discussions on the stability of the ZB
structure, from now on, we like to focus on the electronic and
magnetic properties of the II-V compounds at their equilibrium
positions within the ZB structure.

The electronic structure calculations were performed by using the OpenMX
DFT code \cite{OpenMX }, a linear-combination-of-pseudoatomic-orbitals
method \cite{Ozaki}.  The generalized gradient approximation
(GGA)\cite{Perdew} was used for exchange-correlation potential. The
effects of spin-orbit couplings via a relativistic $j$-dependent
pseudopotential scheme were included in the non-collinear DFT calculations
\cite{MacDonald,Bachelet,Theurich}. Double-valence-plus-single-polarization
orbitals were taken as a basis set, which were generated with cutoff radii
of 6.0, 6.0, and 7.0 {a.u.}  for Ca, Sr, Ba atoms, and 5.0, 5.5, 6.0, 6.5,
and 7.0 {a.u.} for N, P, As, Sb, and Bi atoms, respectively.
Troullier-Martins-type pseudopotentials, with a partial core correction
for all atoms, were used to replace the deep core potentials by
norm-conserving soft potentials in a factorized separable form with
multiple projectors. The real space grid techniques were used with the
energy cutoff of 300 Ry in numerical integrations and the solution of the
Poisson equation using fast Fourier transformations. For the formula cell
containing two atoms, (16$\times$16$\times$16) \textbf{k}-grid was sampled
over the full Brillouin zone. To compare the stability of different
magnetic configurations, we calculated total energies of paramagnetic
(PM), ferromagnetic (FM), and anti-ferromagnetic (AFM) configurations in a
supercell containing 16 atoms with (8$\times$8$\times$8) \textbf{k}-grid.

\begin{figure}
  \begin{center}
    \includegraphics[width=0.48\textwidth]{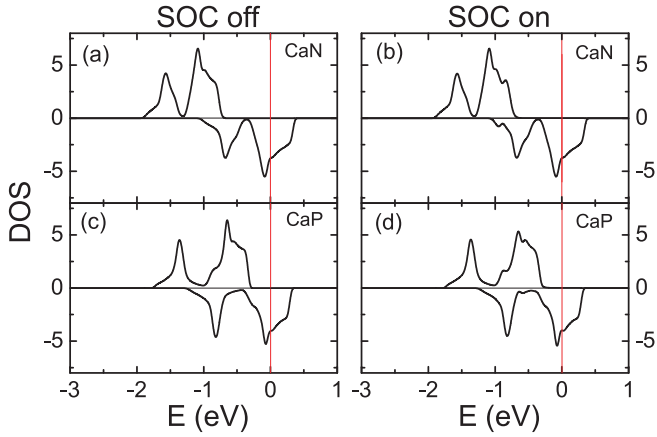}
  \end{center}
  \caption{Total density-of-states of Ca$X$ ($X$ = N and P) without and
    with spin-orbital coupling, as labeled by ``SOC off'' and ``SOC on'',
    respectively.}\label{dosCaNP}
\end{figure}

\begin{figure}
  \begin{center}
    \includegraphics[width=0.48\textwidth]{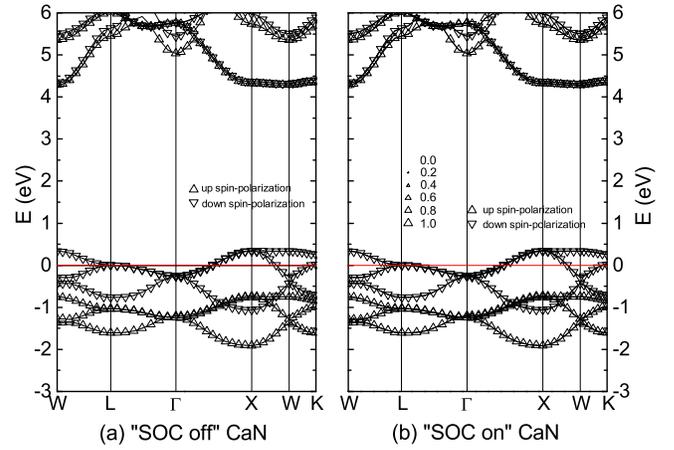}
  \end{center}
  \caption{Band structures of CaN (a) without (``SOC off'') and (b) with
    (``SOC on'') spin-orbital coupling.}\label{bandCaN}
\end{figure}

Figure~\ref{dosCaNP} shows the total density-of-states (DOSs) of ZB CaN
and CaP with and without spin-orbit coupling (SOC). Hereafter we denote
the results with and without SOC by using the labels ``SOC on'' and ``SOC
off'', respectively.  The ``SOC off'' and ``SOC on'' DOS's for both CaN
and CaP are almost identical. It is obvious that the SOC effect is
negligible for the light elements like N or P atoms. Consequently the HMF
character with a magnetic moment of 1 $\mu_{B}$/{f.u.} remains unchanged
regardless of the presence of SOC. In fact, the spin-orbit coupling has
virtually no effect on the ZB CaN band structure, as shown in
Fig.~\ref{bandCaN}.  While the spin-up and spin-down bands are
manifest in the collinear, i.e., ``SOC off'' band structure of
Fig.~\ref{bandCaN}(a), the non-collinear band structure as
obtained from the ``SOC on'' calculations requires a spin-resolved
representation of the bands. In the spin-resolved band plots, as
illustrated in Fig.~\ref{bandCaN}(b) and other ``SOC on'' band
plots in the following figures, we marked each band state by the
symbols ($\bigtriangleup$ or $\bigtriangledown$), the size of
which corresponds to the weight of each spin-component.
Figure~\ref{bandCaN}(b) clearly demonstrates that the
spin-resolved band structure of CaN with negligible SOC has a
clear separation of spin-up and spin-down bands, which is
virtually identical to the collinear bands of
Fig.~\ref{bandCaN}(a).

\begin{figure}
  \begin{center}
    \includegraphics[width=0.48\textwidth]{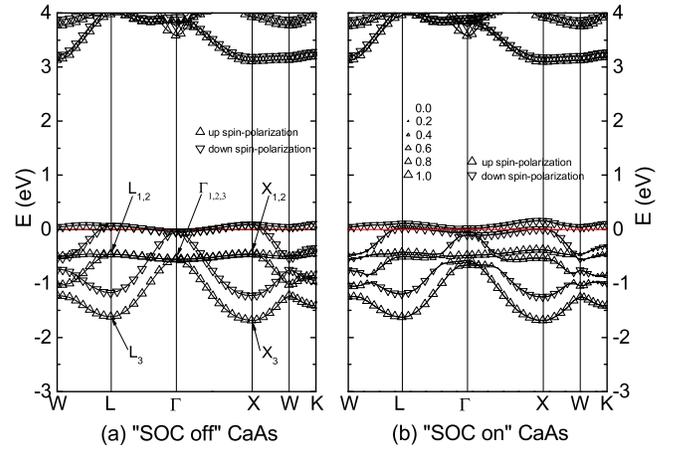}
  \end{center}
  \caption{``SOC off'' and ``SOC on'' band structures of
    CaAs.}\label{bandCaAs}
\end{figure}

\begin{figure}
  \begin{center}
    \includegraphics[width=0.48\textwidth]{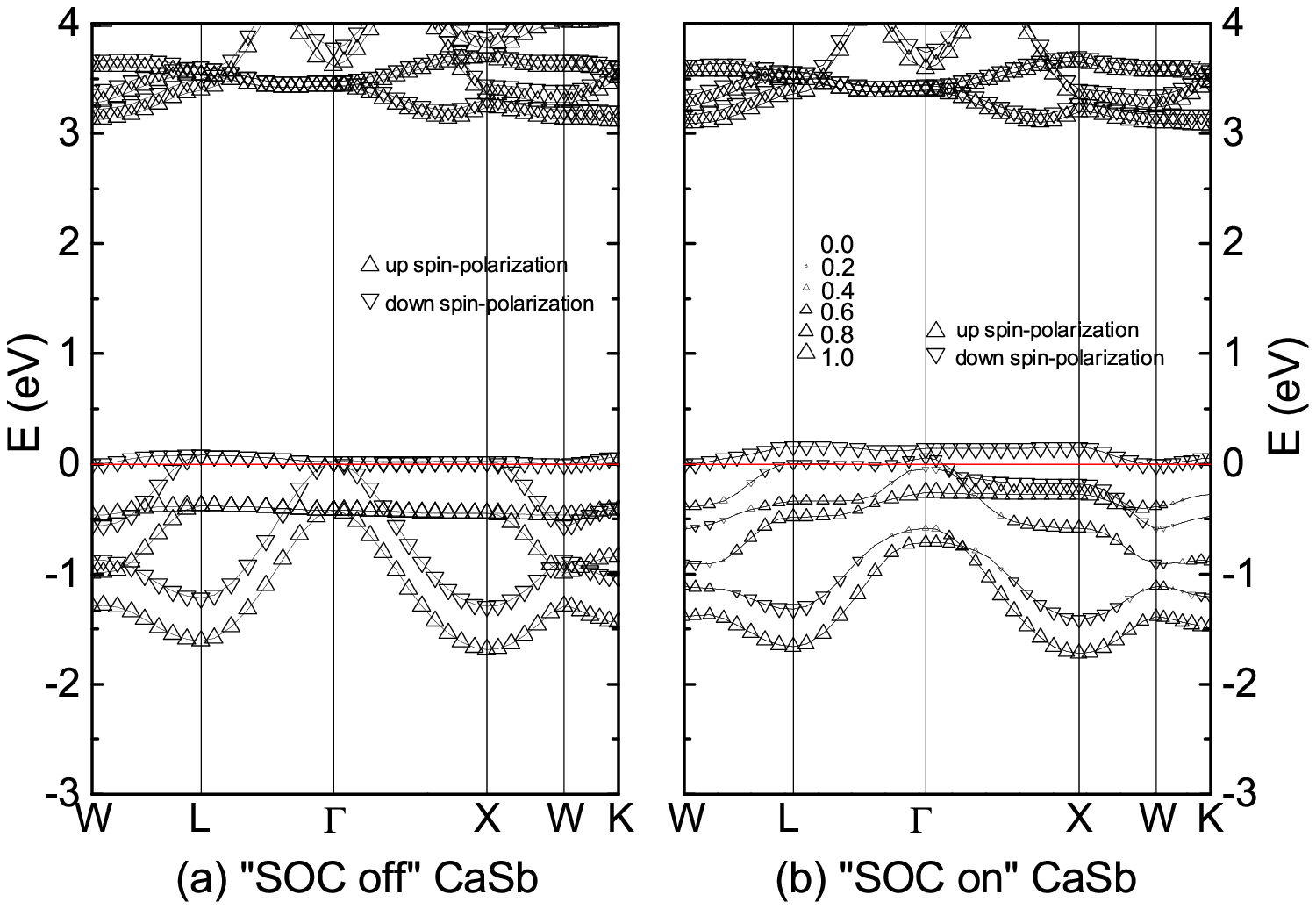}
  \end{center}
  \caption{``SOC off'' and ``SOC on'' band structures of
    CaSb}\label{bandCaSb}
\end{figure}

\begin{figure}
  \begin{center}
    \includegraphics[width=0.48\textwidth]{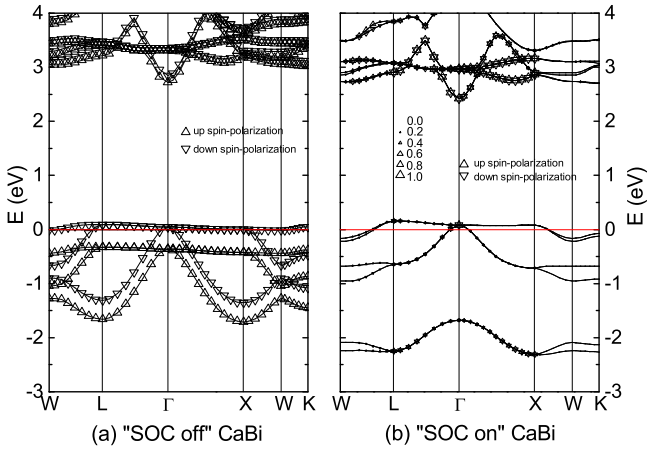}
  \end{center}
  \caption{``SOC off'' and ``SOC on'' band structures of
    CaBi.}\label{bandCaBi}
\end{figure}

To demonstrate the effect of the SOC interactions on the valence band
structures of the ZB II-V compounds, we have calculated the band
structures of Ca$X$ ($X$=As, Sb, and Bi) in addition to CaN and CaP. The
band plots of CaAs, CaSb, and CaBi are shown in Figs.~\ref{bandCaAs},
\ref{bandCaSb}, and \ref{bandCaBi}, respectively. One of the common
features of the ``SOC off'' band structures of Ca$X$ is a strong
exchange-splitting of the flat $p$ bands consisting of the anion
$p$ states. The origin of magnetism for the HMF Ca$X$ compounds was
attributed to the magnetic instability of the narrow $p$ bands at the
Fermi level\cite{II-Sie}. In fact, the flatness of the $p$ bands are more
pronounced in the ``SOC off'' band structures of Ca$X$ ($X$=As, Sb, and
Bi) than those of CaN and CaP.

The nature of the flat $p$ bands in CaAs has been discussed in the
previous work\cite{II-Sie} based on the tight-binding model. We can
understand the valence band structure of $AX$ by a simple fcc lattice of
anion $X$ ($X$= N, P, As, Sb, and Bi) with an extraordinarily large
lattice constant. In this simplified picture, for instance, the valence
band structure of $AX$ can be mimicked by the partially filled $p$ bands
of a fcc Br solid, which features the same $p$ band with the same electron
filling as that of $AX$ except the flat bands observed in $AX$. Indeed, to
acquire the flatness, the $X$ $p$-$p$ hybridization should be
counter-balanced by the $A$ $d$-$X$ $p$ hybridization.

\begin{figure}
  \begin{center}
    \includegraphics[width=0.48\textwidth]{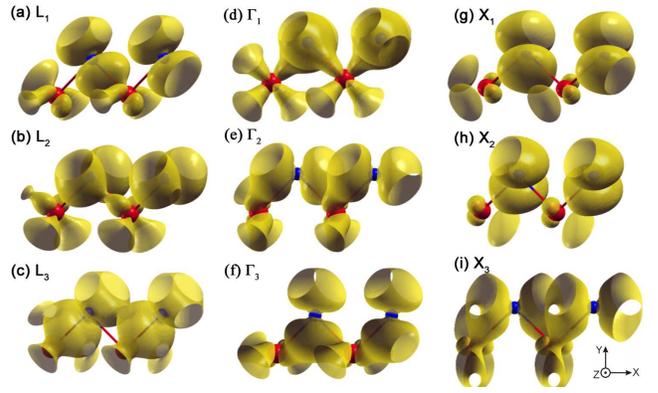}
  \end{center}
  \caption{(Color online) Charge density profiles of (a) $L_{1}$, (b)
    $L_{2}$, (c) $L_{3}$, (d) $\Gamma_{1}$, (e) $\Gamma_{2}$, (f)
    $\Gamma_{3}$, (g) $X_{1}$, (h) $X_{2}$, and (i) $X_{3}$ states of the
    ``SOC off'' CaAs bands, which are labelled in
    Fig.~\ref{bandCaAs}(a). Light (red) spheres represent for Ca atoms and
    blue (dark) spheres for As atoms.}\label{chgCaAs}
\end{figure}

Figure~\ref{chgCaAs} illustrates charge density profiles of the ``SOC
off'' spin-up states of CaAs, which are labelled by $\Gamma_{i}$,
$L_{i}$, and $X_{i}$ ($i$=1,2,3) in Fig.~\ref{bandCaAs}(a). At the
$\Gamma$ point, the triply degenerate $\Gamma_{i}$ ($i$=1,2,3) states
consist of $p_{x}$, $p_{y}$, and $p_{z}$ orbitals and form a anti-bonding
configuration among the As $p$ states. Due to the presence of Ca, however,
the $p$-$p$ anti-bonding states are mixed with the Ca $d$-As $p$ bonding
configurations. Without the Ca $d$-As $p$ bonding contribution, the energy
level of the $\Gamma_{i}$ states should have lied at the higher position
than those shown in Fig.~\ref{bandCaAs}(a). As the $\mathbf{k}$ vector
moves along $\Gamma X$ and $\Gamma L$, the $L_{1,2}$ and $X_{1,2}$ states
change into a non-bonding configuration, while the $L_{3}$ and $X_{3}$
states develop into a strong As $p$-As $p$ bonding configuration,
respectively, as illustrated in Fig.~\ref{chgCaAs}(c) and (i).  Since the
nearly flat bands of the doubly degenerate
$L_{1,2}\!-\!\Gamma_{1,2}\!-\!X_{1,2}$ states contribute to the high DOS at
$E_{\mathrm{F}}$, when the SOC is turned off, the Stoner instability would
drive the system into the ferromagnetic ground state.


Now let us consider the effect of spin-orbit coupling interactions on the
valence band structure. When SOC is turned on, contrary to the robust HMF
band structures of CaN and CaP, the electronic structures of Ca$X$
($X$=As, Sb, and Bi) change remarkably.  Despite that the ``SOC off''
valence bands of all the Ca$X$ ($X$=As, Sb, and Bi) compounds are almost
identical as shown in Fig.~\ref{bandCaAs}--\ref{bandCaBi}, the band
dispersion and spin characters of the ``SOC on'' band structures are
markedly different and evolve as the atomic number of the element $X$
increases from As to Sb to Bi.  Since the SOC strength becomes larger for
the heavier atoms, the spin-resolved band structures exhibit more
complicated features. Before discussing the details of the SOC effects on
the spin-resolved band structures, let us examine the evolution of the
``SOC on'' DOS of Ca$X$ ($X$=As, Sb, and Bi).

\begin{figure}
  \begin{center}
    \includegraphics[width=0.48\textwidth]{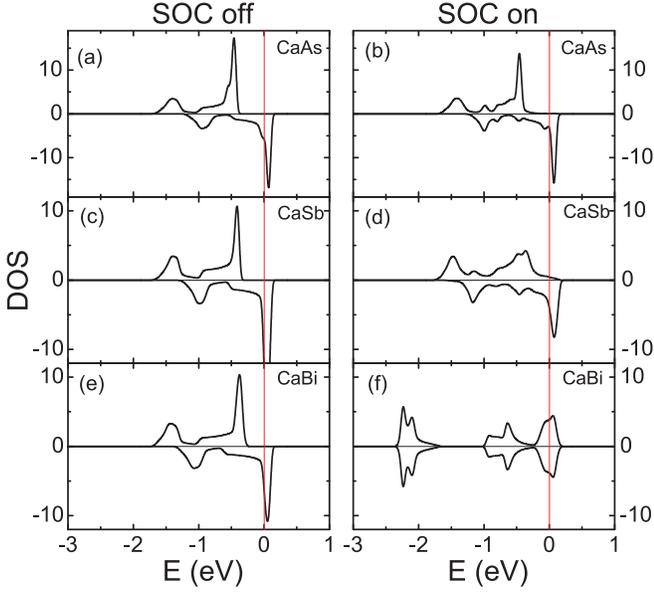}
  \end{center}
  \caption{Total density-of-states of Ca$X$ ($X$=As, Sb, and Bi) without
    (``SOC off'') and with (``SOC on'') spin-orbital
    coupling.}\label{dosCaAsSbBi}
\end{figure}

The ``SOC off'' DOSs of Ca$X$ ($X$=As, Sb, and Bi) in
Fig.~\ref{dosCaAsSbBi}(a), (c) and (e) have basically the same
characteristics as those of CaN and CaP in Fig.~\ref{dosCaNP}(a) and (c)
except the pronounced flatness in the Ca$X$ DOSs. As the SOC strength
increases from CaAs to CaSb, the sharp peaks corresponding to the flat
$p$ bands broaden and the spin-up and down components seem to smear into
each other. (See Fig.~\ref{dosCaAsSbBi}(d) of CaSb.) For the CaBi case,
where the SOC energy scale is dominant, the DOS feature shows no
spin-polarization but a clear separation of two peaks: one centered at
$-2.2$ eV corresponding to the $j$=1/2 state and the other at $-0.5$ eV to
$j$=3/2, as shown in Fig.~\ref{dosCaAsSbBi}(f) of CaBi.

In addition to the evolution of the DOS features, one can observe the
change of spin-polarization (SP) at $E_{\mathrm{F}}$ as well as the
magnetic moment ($M$). The change of SP can be easily understood from the
``SOC on'' DOS plots of Fig.~\ref{dosCaAsSbBi}. The spin-polarization of
CaAs is close to but not exactly 100\%, which means that CaAs is not a HMF
in a strict sense, whereas the spin-polarization of CaBi is zero.  While
the half-metallicity in CaAs and CaSb is destroyed moderately, the
half-metallic character of CaBi disappears completely. 

\begin{table}
  \caption{\label{EleMag}
    Spin-polarization (SP) at Fermi level and magnetic moment ($M$)
    $\mu_{B}$/f.u. as obtained from the ``SOC on'' calculations.}
  \begin{tabular}{cccccc} \hline\hline
    &N       &P       &As      &Sb      &Bi \\
    & SP / $M$   & SP / $M$   & SP / $M$   & SP / $M$   & SP / $M$ \\ \hline
    Ca      &1.00 / 1.00   &1.00 / 1.00   &0.96 / 1.10    &0.81 / 1.11    &0 / 0\\
    Sr      &1.00 / 1.00   &1.00 / 1.00   &0.97 / 1.10    &0.84 / 1.17    &0 / 0\\
    Ba      &1.00 / 1.00   &1.00 / 1.00   &0.97 / 1.03    &0.79 / 1.00    &0 / 0\\
    \hline\hline
  \end{tabular}
\end{table}

Table~\ref{EleMag} summarizes the spin polarization (SP) at
$E_{\mathrm{F}}$ and the magnetic moment ($M$) of $AX$ ($A$=Ca, Sr, and
Ba; $X$=N, P, As, Sb, and Bi). Here the SP is defined by the ratio of the
spin-up and spin-down DOS at $E_{\mathrm{F}}$ over the total DOS at
$E_{\mathrm{F}}$. The magnetic moments in Table~\ref{EleMag} include the
contribution from both the spin and orbital moments as obtained from the
non-collinear calculations. Due to the smearing of the spin-up and down
components near $E_{\mathrm{F}}$, both As and Sb compounds have
non-negligible contribution of the majority spin DOS at
$E_{\mathrm{F}}$. The mixed spin-components in the valence bands are
clearly shown in the spin-resolved band structures of
Fig.~\ref{bandCaAs}(b) and Fig.~\ref{bandCaSb}(b). As the SOC strength
increases, some of the non-collinear band states have a strong mixture of
spin-up and down components.

The ``SOC on'' band structures of Ca$X$ ($X$=N, P, As, Sb, and Bi)
shown in Fig.~\ref{bandCaN}--\ref{bandCaBi} can be classified into
three categories: (i) the CaN-type where the spin-exchange energy
of $\Delta E_{ex}\approx 0.5$ eV is dominant, (ii) the CaBi-type
where $j$=1/2 and $j$=3/2 bands are well separated by the SOC
energy scale of $\Delta E_{so}\approx 2$ eV, and (iii) the
CaSb-type, an intermediate case where both $\Delta E_{ex}$ and
$\Delta E_{so}$ compete with each other. In the case of (i) the
CaN-type, the SOC interactions are negligible, i.e., $\Delta
E_{ex}\gg \Delta E_{so}$, so that the ferromagnetic band structure
is set by the exchange instability triggered by the narrow $p$
bands. On the other hands, to understand the band structures of
the types (ii) and (iii), one has to consider both SOC and
exchange interactions on the same basis. To estimate the
magnitudes of $\Delta E_{ex}$ and $\Delta E_{so}$, we calculated
the exchange and spin-orbit splitting energies at the
$\Gamma$-point for the ``SOC off'' ferromagnetic and ``SOC on''
paramagnetic states, respectively. The calculated values of $\Delta
E_{ex}$ and $\Delta E_{so}$ for $AX$ ($A$=Ca, Sr, and Ba; $X$=As, Sb, and
Bi) are listed in Table~\ref{ExcEso}.

\begin{table}
  \caption{\label{ExcEso} Exchange and spin-orbit splitting
  energies in $AX$ ($A$=Ca, Sr, and Ba; $X$=As, Sb, and Bi). Energies are
in eV unit.}

  \begin{tabular}{cccc} \hline\hline
         &As            &Sb            &Bi \\
      & $\Delta E_{ex}$ / $\Delta E_{so}$ & $\Delta E_{ex}$ / $\Delta
      E_{so}$  & $\Delta E_{ex}$ / $\Delta E_{so}$  \\ 
    \hline
    Ca      &0.51 / 0.25     &0.44 / 0.52     &0.41 / 1.88    \\
    Sr      &0.51 / 0.23     &0.44 / 0.52     &0.41 / 1.77    \\
    Ba      &0.48 / 0.21     &0.41 / 0.47     &0.43 / 1.62    \\
    \hline\hline
  \end{tabular}
\end{table}

For the case of (ii) the CaBi-type, the SOC energy of $\Delta E_{so}=1.88$
eV is much larger than the effective exchange energy of $\Delta{}
E_{ex}=0.41$ eV. Since the SOC term dominates, the valence band
eigenstates of CaBi at the $\Gamma$-point are well represented by $j$=3/2
and $j$=1/2 states, corresponding to the 4-fold and 2-fold degeneracy,
respectively. Even away from the $\Gamma$-point, the spin-up and down
components are mixed up by maintaining its character of the spin-orbit
coupled $j$-states. Consequently the average spin-polarization in CaBi
becomes null so that the ground state is non-magnetic.

The valence band structures of CaSb, SrSb, and BaSb are much more
complicated as the energy scales of the SOC and exchange interactions
become comparable to each other. For an example, as listed in
Table~\ref{ExcEso}, the SOC and exchange energies for CaSb are $\Delta
E_{so}=0.52$ eV and $\Delta E_{ex}= 0.44$ eV, respectively. The overall
shape of the ``SOC on'' valence bands of CaSb in Fig.~\ref{bandCaSb}(b)
resembles those of the ``SOC off'' valence bands in
Fig.~\ref{bandCaSb}(a), whereas the degeneracy of the bands along
$L\!-\!\Gamma\!-\!X$ is significantly destroyed. Due to the strong
contribution of $\Delta E_{so}$, the spin-flip term
($L_{+}S_{-}\!+\!L_{-}S_{+}$) in the SOC term of $\mathcal{H}_{so} =
\lambda_{so} \mathbf{L}\cdot\mathbf{S}$ acts as a perturbation to the
exchange-split states.


In conclusion, we have shown that the half-metallic ferromagnetism in ZB
II-V compounds can be destroyed for $AX$ ($A$=Ca, Sr, and Ba; $X$=As, Sb,
and Bi) due to the presence of strong spin-orbit coupling
interactions. Our non-collinear DFT calculations revealed the evolving
features of the spin-orbit coupling and exchange-split in the
spin-resolved band structures as the atomic number increases from As to Sb
to Bi. Considering the variation of the spin polarizations of ZB II-V
compounds, it is suggested that one can control the SOC strength to design
either HMF or paramagnetic metal. This SOC-dependent feature may be useful
in the spintronics application of HMF materials.


We are grateful to Prof.\ J.~I.~Lee for helpful suggestions and
discussions. This work was supported by the KRF Grant (MOEHRD
KRF-2005-070-C00041). We also acknowledge the support from KISTI (Korea
Institute of Science and Technology Information) under The Supercomputing
Application Focus Support Program.

\end{document}